# A Casual Tour Around a Circuit Complexity Bound[*]

Ryan Williams[†]

**Abstract**

I will discuss the recent proof that the complexity class NEXP (nondeterministic exponential time) lacks nonuniform ACC circuits of polynomial size. The proof will be described from the perspective of someone trying to discover it.

## 1  Background

Recently, a new type of circuit size lower bound was proved [Wil10, Wil11]. The proof is a formal arrangement of many pieces. This article will not present the proof. It will not present a series of technical lemmas, each lemma following from careful logical arguments involving the previous ones, culminating in the final result. This article is a discussion about how to discover the proof – a casual tour around it. Not all details will be given, but you will see where all the pieces came from, and how they fit together. The path will be littered with my own biased intuitions about complexity theory – what I think should and shouldn't be true, and why. Much of this intuition may well be wrong; however I can say it has led me in a productive direction on at least one occasion. I hope this article will stimulate the reader to think more about proving lower bounds in complexity.

The remainder of this section briefly recalls some (but not all of the) basics that we'll use.

### 1.1  Uniform Complexity: Algorithms

Recall that $\mathsf{NTIME}[t(n)]$ denotes the class of languages (decision problems) solvable by nondeterministic algorithms running in time $t(n)$ on inputs of length $n$. So $L \in \mathsf{NTIME}[t(n)]$ provided that there is a nondeterministic algorithm such that for all strings $x \in L$, there is a computation path of $t(|x|)$ steps that results in acceptance of $x$, and for $x \notin L$, every possible computation path of $t(|x|)$ steps results in rejection of $x$. We define $\mathsf{NEXP} = \bigcup_{k>0} \mathsf{NTIME}[2^{n^k}]$. NEXP informally corresponds to problems with exponentially long solutions which are verifiable in exponential time. This class encompasses everything considered feasibly computable (and more): NP, PSPACE, EXP, etc.

The *nondeterministic time hierarchy* [SFM78, Zak83] says that, as one permits longer solutions to problems and longer verification time for those solutions, one can always find strictly more problems with verifiable solutions, under the constraints. In notation, $\mathsf{NTIME}[t(n)] \subsetneq \mathsf{NTIME}[T(n)]$ for $t(n+1) \leq o(T(n))$. One consequence of the nondeterministic time hierarchy is that $\mathsf{NEXP} \neq \mathsf{NP}$.[1]

---

[*]An earlier version of this article appears in *SIGACT News*, September 2011.

[†]Computer Science Department, Stanford University, Stanford, CA, USA. rrwilliams@gmail.com. At the time of writing, the author was supported by the Josef Raviv Memorial Fellowship at IBM Research – Almaden.

[1]Actually, the usual deterministic time hierarchy suffices to prove this, using a padding/translation argument.

## 1.2 Nonuniform Complexity: Circuits

With the usual uniform models of computation (Turing machines, lambda calculus, $\mu$-recursive functions, etc.), a function only counts as *computable* provided we can find a single algorithm in the model that computes the function on all possible finite inputs. Similarly, in complexity theory, a function is only *efficiently computable* if a single algorithm runs efficiently on all finite inputs.

Suppose I allow you to run a different algorithm $A_n$ for every distinct input length $n$. This amounts to having a countably infinite set of algorithms, which looks unrealistic.[2] But by permitting the length of a program to grow with the input length, we can more accurately model algorithms in practice that exploit the fact that there is an upper bound on the inputs they receive. Could there be a program for 3SAT with a billion lines of code that rapidly solves 3SAT on all formulas with less than a billion clauses? Nonuniform complexity can address questions of this form.

We will imagine an infinite family of algorithms $\{A_n\}$ as a family of logical circuits, where $A_n$ takes $n$ bits of input and returns a bit. (We'll work exclusively over the binary alphabet, for simplicity.) The classes of circuits considered in this article are, in increasing order of expressiveness:

- $\mathsf{AC}^0$, the class of circuits with constant depth and polynomial size, having unbounded fan-in AND, OR, and NOT gates,

- $\mathsf{AC}^0[m]$, the class of circuits with constant depth and polynomial size, having unbounded fan-in MOD$m$, AND, OR, and NOT gates (where a MOD$m$ gate outputs 1 iff the sum of its inputs is divisible by $m$),

- ACC, the union over all $m$ of the classes $\mathsf{AC}^0[m]$,[3] and

- $\mathsf{P}/\mathsf{poly}$, the class consisting of arbitrary polynomial size Boolean circuits with bounded fan-in AND and OR gates, and NOT gates.

(One can take the "size" of a circuit to be either the number of gates or the number of wires; for us, the choice won't matter.) We will identify the circuit classes above with their corresponding *language classes*, which consists of all decision problems solvable with an infinite family of such circuits. So, "$\mathsf{NP} \subset \mathsf{P}/\mathsf{poly}$" states that every problem in NP can be solved with an infinite circuit family $\{C_n\}$ drawn from $\mathsf{P}/\mathsf{poly}$, where each $C_n$ is run on inputs of length $n$.

A routine fact is that $\mathsf{P} \subset \mathsf{P}/\mathsf{poly}$: problems solvable in *polynomial time* by some algorithm can be solved by an infinite family of polynomial size circuits. Therefore, our restriction to considering circuits rather than arbitrary "growing" programs is actually without loss of generality: a poly($n$)-time program with poly($n$) lines of code can be simulated by a poly($n$)-size circuit on all $n$-bit inputs, although the underlying polynomials may not have the same degrees.[4]

---

[2]Indeed, one can solve undecidable problems using an infinite number of algorithms: let $A_n(1^n)$ output 1 iff the $n$th Turing machine halts on blank tape. This is a fact that is either true or false for a given $n$, so for each $n$ we can make a very short efficient program $A_n$ that captures this behavior.

[3]Note that an ACC circuit family doesn't necessarily have to be polynomial sized. By default, a circuit's size should be assumed polynomial unless otherwise specified.

[4]We use the notation poly($n$) to denote expressions of the form $n^k$ for some fixed $k$ independent of $n$.

## 1.3 Ruling Out Polynomial Size Circuits for Uniform Complexity Classes

What uniform computations can be simulated in P/poly? This is largely open. Randomized complexity classes like RP and BPP are in P/poly, but we do not believe that NP-complete problems can be solved with polynomial size circuit families. However, proving NP $\not\subset$ P/poly is only stronger than P $\neq$ NP. The "smallest" uniform complexity class that we know is not contained in P/poly is $MA_{EXP}$, the exponential-time version of Merlin-Arthur games [BFT98]. Kabanets and Impagliazzo [KI04] "almost" proved that the slightly smaller class $NEXP^{RP}$ (nondeterministic exponential time with an RP oracle) isn't in P/poly: either $NEXP^{RP}$ doesn't have *arithmetic* polysize circuits, or it doesn't have (the usual Boolean) polysize circuits. Both $MA_{EXP}$ and $NEXP^{RP}$ are enormous classes, containing NEXP and more.

So while we can show that certain functions cannot have polynomial size circuits, those functions are *extremely* difficult for uniform algorithms to compute. But it could still be true that $EXP^{NP} \subset$ P/poly. This looks crazy; if true, it would not only mean that every problem with *exponentially long solutions* can be solved with *polynomial size circuits*, but that every problem in $EXP^{NP}$ has "highly compressible" solutions, representable with polynomial size circuits! Moreover, we will see in Section 5 that a similar result holds assuming NEXP $\subset$ P/poly.

## 1.4 ACC: The Frontier

Can we make progress on P/poly lower bounds, by considering more restricted classes of circuits? In the space of this article we can only give a condensed history; much more can be found in the surveys [All96, Vio09]. Furst, Saxe, and Sipser [FSS81] and Ajtai [Ajt83] proved that simple functions such as the parity of $n$ bits cannot be computed by polynomial size $AC^0$ circuits. (These results were later strengthened to exponential size [Yao85, Hås86].) A natural next step was to grant $AC^0$ the parity function for free – resulting in the study of $AC^0[2]$. Razborov [Raz87] proved an exponential size lower bound for computing the majority of $n$ bits in $AC^0[2]$. Smolensky [Smo87] proved exponential lower bounds for computing $MOD_q$ with $AC^0[p]$, for distinct primes $p$ and $q$. Barrington [Bar89] suggested the next step would be to prove lower bounds for the class ACC which allows for $MOD_m$ gates where $m$ can be an arbitrary constant.

Although it was conjectured over 20 years ago that the majority of $n$ bits cannot be computed with ACC, strong ACC lower bounds have escaped proof. Suppose we grant $AC^0$ both the $MOD_2$ and the $MOD_3$ function for free; this is equivalent to studying $AC^0[6]$, as seen by the equations

$$\begin{aligned} MOD_6(x_1,\ldots,x_n) &= MOD_3(x_1,\ldots,x_n) \wedge MOD_2(x_1,\ldots,x_n), \\ MOD_2(x_1,\ldots,x_n) &= MOD_6(x_1,x_1,x_1,\ldots,x_n,x_n,x_n), \\ MOD_3(x_1,\ldots,x_n) &= MOD_6(x_1,x_1,\ldots,x_n,x_n). \end{aligned}$$

Even for this class, it was still possible that $EXP^{NP} \subset AC^0[6]$! Given that $AC^0[p]$ was known to be very weak for every prime $p$, this was an extremely frustrating open problem – how could $MOD_6$ be so much more powerful than $MOD_7$?

The recent paper [Wil11] finally rules out this ludicrous possibility. For example, we can prove that $AC^0[6]$ circuits for $EXP^{NP}$ must necessarily have at least $2^{n^\delta}$ size, for some $\delta > 0$ that depends on the circuit depth. The proof extends to the (smaller) class NEXP as well, although there is some loss in the size lower bound. Nevertheless we can still rule out polynomial size ACC circuits for NEXP (even quasipolynomial size). The basic framework behind the proof is generic enough that it is reasonable to believe it can be extended to prove much stronger results: perhaps NEXP $\not\subset$ P/poly, or NP $\not\subset$ ACC, or more.

## 2 Acquiring The Target

Suppose we've set ourselves to finding a problem in NEXP that cannot be in ACC. What is a good NEXP problem to choose? The "hardest" possible candidates should be NEXP-complete ones – if there's a problem in NEXP\ACC, then the complete ones are there! However, NEXP-completeness hasn't been studied nearly as much as NP-completeness, so the list of NEXP-complete problems doesn't appear to be terribly long. Nevertheless there is a natural way to construct NEXP problems out of NP problems, by focusing on the highly compressible instances of NP-complete problems.

Given a problem $\Pi$, we define the SUCCINCT $\Pi$ problem as follows. Let $\mathcal{C}$ be the set of all Boolean circuits with a single output gate over the gate basis AND/OR/NOT. For every $C \in \mathcal{C}$, let $T(C)$ be the *truth table* of the function represented by $C$. More formally, letting $n$ be the number of inputs to $C$, $T(C)$ is the $2^n$ bit string where $T(C)[i] = C(s_i)$, where $s_i$ is the $i$th $n$-bit string (in lexicographical order, say).

**Problem:** SUCCINCT $\Pi$
*Given*: A circuit $C$ from $\mathcal{C}$ with $n$ inputs and poly($n$) size.
*Task:* Determine whether $T(C)$ is a yes-instance of $\Pi$, i.e., $T(C) \in \Pi$.

So in SUCCINCT $\Pi$, we only wish to solve the "highly compressible" instances of $\Pi$: those $2^n$ bit instances which are compressible to poly($n$)-bit representations as circuits.

The definition may look odd at first, but studying succinct problems is something that many of us already do. Consider the *OR problem*: given a bit string $x$, does $x$ contain a 1? This problem is trivial from the time complexity perspective, but still interesting on the circuit complexity level, as it is not known whether constant-depth circuits made entirely of $MOD_6$ gates can compute OR efficiently [HK09]. However, the SUCCINCT-OR problem is exactly the NP-complete Circuit Satisfiability problem: given a circuit, does its truth table contain a 1? So even the succinct versions of trivial problems are already interesting.[5]

What if $\Pi$ is an NP-complete problem? How hard is SUCCINCT $\Pi$? There are nondeterministic exponential time algorithms for solving such problems:

**Proposition 1** *Let $\Pi$ be a NP-complete problem that admits proofs of length $\ell(n)$ for n-bit instances, with a verifier that runs in $t(\ell)$ time on proofs of length $\ell$. SUCCINCT $\Pi$ can be solved in $t(\ell(2^n)) + 2^n \cdot poly(s)$ nondeterministic time, on circuits of size s with n inputs.*

**Proof.** Evaluate the given circuit on all of its possible inputs in $2^n \cdot \text{poly}(s)$ time, producing an instance of $\Pi$ of length $2^n$. By assumption, the instance has a proof (if one exists) of length $\ell(2^n)$, and the proof can be verified in $t(\ell(2^n))$. Nondeterministically guessing the $\ell(2^n)$-bit proof and verifying that proof yields the running time. □

As an example, consider the succinct version of 3SAT:

**Corollary 2.1** SUCCINCT 3SAT *can be solved in nondeterministic $2^n \cdot (poly(s) + poly(n))$ time on circuits of n inputs and s size.*

---
[5]Encyclopedias could be written on succinct representations of problems in computer science. In complexity theory, succinctly represented problems are closely related to the structural notion of sparse sets; this is best illustrated by Hartmanis, Immerman, and Sewelson's theorem that $\mathsf{TIME}[2^{O(n)}] \neq \mathsf{NTIME}[2^{O(n)}]$ iff there is a sparse set in NP\P [HIS85]. Implicit representations of graphs have been widely studied, and solving problems on them amounts to solving the succinct version of a graph problem. In other communities, BDDs (Binary Decision Diagrams) are the standard means for representing functions; many problems studied in that arena can be seen as succinct problems where the underlying circuit class $\mathcal{C}$ has been replaced with the set of BDDs. The Wikipedia articles on these (particular) topics are good starting points for further references.

**Proof.** Apply the above proposition. Here the proofs are satisfying assignments, which do not exceed the length of a formula, so $\ell(n) \leq n$. Verifying a satisfying assignment for a $2^n$-size 3-CNF can be done in $O(2^n \cdot \text{poly}(n))$ time. □

Papadimitriou and Yannakakis [PY86] showed that for all known NP-complete problems $\Pi$, SUCCINCT $\Pi$ is NEXP-complete. We state their result informally:

**Theorem 2.1 ([PY86])** *If $\Pi$ is NP-complete under "ultra-efficient reductions" then* SUCCINCT $\Pi$ *is NEXP-complete.*

Essentially what is needed in an "ultra-efficient reduction" is that each bit of the reduction's output can be computed from a polylogarithmic number of bits of the input, in polylogarithmic time. Now we have our pick of candidate NEXP problems: the succinct versions of NP-complete problems are fair game.

What NP-complete problem could be more natural than 3SAT? It has been studied to death; the literature is filled with theorems on it. An attractive property of SUCCINCT 3SAT is that it's *very* NEXP-complete: there are super-ultra-efficient reductions from arbitrary languages in NEXP to SUCCINCT 3SAT instances. So there is little loss of generality in focusing on SUCCINCT 3SAT.

**Theorem 2.2 (Efficient Cook-Levin for NEXP)** SUCCINCT 3SAT *is NEXP-complete under polynomial time reductions. Moreover, there is a polynomial time reduction $R$ from arbitrary $L \in \text{NTIME}[2^n]$ to* SUCCINCT 3SAT *with the properties:*

- *$x \in L \iff R(x) \in$ SUCCINCT 3SAT; i.e., $R(x)$ is a circuit such that $T(R(x))$ encodes a satisfiable 3-CNF formula.*

- *$R(x)$ is a circuit with poly($|x|$) gates.*

- *For all sufficiently long $x$, the number of inputs to the circuit $R(x)$ is at most $|x| + 4\log|x|$.*

The first two properties could be met rather straightforwardly, if the circuit $R(x)$ were allowed to have up to $O(|x|)$ inputs. One of the many textbook proofs that 3SAT is NP-complete would suffice. We can convert a nondeterministic time $t$ computation $A(x)$ into a $t^{c+1}$ size 3-CNF formula, by first translating the computation of $A(x)$ into a nondeterministic one-tape Turing machine $M(x)$ running in time $t^c$ and using space $t$, then building a $t^c \times t$ matrix $T_x$ where $T_x(i,j)$ holds the content of the $j$th cell of $M(x)$ at step $i$ of its execution. (And if the head is reading cell $j$ at step $i$, then $T_x(i,j)$ also holds the state of $M(x)$ at step $i$.) Note $T_x$ is often called a *tableau*. (The particular value of $c$ depends on the original computational model: if the model is multitape Turing machines, then $c = 2$ suffices.)

Observing that every entry in $T_x$ can be determined from at most three other entries, we can generate constant size 3-CNF formulas, one for each entry of $T_x$, such that their conjunction is satisfiable if and only if $A(x)$ accepts. This 3-CNF formula generated is extremely regular, in that essentially the same group of clauses is produced repeatedly (with only minor changes in the variable indices). It follows that the clauses corresponding to entry $T(i,j)$ can be efficiently produced with a poly($\log t$)-size circuit that is given $(i,j) \in [t^c] \times [t]$ as a $(c+1)\log t + O(1)$-bit string. When $t = 2^n$, we obtain a poly($n$)-size circuit with about $(c+1)n$ inputs.

However, more efficient proofs of the Cook-Levin theorem exist, and the formulas obtained there have high redundancy too. Even for random access machines, there is a reduction from time-$t$ computation to $O(t\log^4 t)$ size formulas where the $i$th bit of the formula can be computed (given the integer $i$ as an input) in

poly$(\log t)$ time [Coo88, Rob91, FLvMV05]. This corresponds to a reduction in the number of inputs to the circuit $R(x)$, from $(c+1)|x|$ down to $|x|+4\log|x|$. There are several ways to achieve this kind of reduction, but unfortunately we do not have the space to include intuition for them; please consult the references above.

We saw earlier that SUCCINCT 3SAT can be solved nondeterministically in $2^n s^{O(1)}$ time, on circuits with $s$ gates and $n$ inputs. Theorem 2.2 implies a time *lower bound* on how efficiently SUCCINCT 3SAT can be solved nondeterministically.

**Theorem 2.3 (Time Lower Bound for** SUCCINCT 3SAT**)** SUCCINCT 3SAT *cannot be solved in* $2^{n-\omega(\log n)}$ *time (even with nondeterminism) on circuits with n inputs and* poly$(n)$ *gates.*

**Proof.** Assume SUCCINCT 3SAT had a nondeterministic algorithm with the above running time. By the Cook-Levin Theorem for NEXP (Theorem 2.2), every $n$-bit instance of every $L \in \mathsf{NTIME}[2^n]$ can be reduced in poly$(n)$ time to a SUCCINCT 3SAT circuit $C$ with $n+4\log n$ inputs and poly$(n)$ size. By assumption, the "succinct satisfiability" of $C$ can be determined in $2^{(n+O(\log n))-\omega(\log n)}$poly$(n) \leq o(2^n)$ time, with a nondeterministic algorithm. Therefore every $L \in \mathsf{NTIME}[2^n]$ is contained in the class $\mathsf{NTIME}[o(2^n)]$, i.e., $\mathsf{NTIME}[2^n] \subseteq \mathsf{NTIME}[o(2^n)]$. But this contradicts the nondeterministic time hierarchy theorem [SFM78, Zak83] which says $\mathsf{NTIME}[o(2^n)] \subsetneq \mathsf{NTIME}[2^n]$. □

So there is a concrete limitation on how efficiently SUCCINCT 3SAT can be solved, and it looks pretty strong. Could this result on the time complexity of SUCCINCT 3SAT be translated into a limitation on the circuit complexity? Let's think back to why we believe that separations like $\mathsf{NEXP} \not\subset \mathsf{ACC}$ are true. We believe that problems in nondeterministic exponential time cannot be solved with polynomial size circuits, simply because exponentials grow much faster than polynomials. This is the main reason why we can diagonalize and prove $\mathsf{NEXP} \neq \mathsf{NP}$, but this observation is not at all enough to prove a *nonuniform* lower bound against NEXP. We have to show that even if one were allowed infinite time to rig up infinitely many polynomial size circuits, each devoted to a separate input length $n$, one still cannot solve SUCCINCT 3SAT with this model. The diagonalization argument used in the proof of $\mathsf{NEXP} \neq \mathsf{NP}$ won't work here, and this is "provably" true. (More formally, the diagonalization argument works relative to every oracle, but there are oracles relative to which $\mathsf{NEXP} \subset \mathsf{P}/\mathsf{poly}$.)

Still, it is hard to let go of a strong feeling that polynomial size circuits simply contain too little information to carry out a full simulation of an exponential time computation. Although you are given a separate circuit for each input length, that little circuit is completely representative of some time-intensive function's behavior on an exponential number of inputs. A polynomial size circuit for a function means that the function's truth table is highly compressible and regular. In that sense, polynomial size circuits seem much closer to *polynomial time algorithms* than to exponential time algorithms. We'd like to say that, if there were small circuit families for a problem like SUCCINCT 3SAT, then there may as well be time efficient *algorithms* for SUCCINCT 3SAT. That is, if SUCCINCT 3SAT had polynomial size circuits, then these "short representatives" of exponential time computation may be discovered algorithmically in an efficient way. More generally, the mere existence of these short representatives should mean that SUCCINCT 3SAT has so much problem structure that this structure can also be exploited algorithmically.

At this point we have reached a degree of handwaving so exuberant, one may fear we are about to fly away. Surprisingly, this handwaving has a completely formal theorem behind it:

**Theorem 2.4 (Spinning Circuits Into Algorithms [Wil11])** *If* SUCCINCT 3SAT *can be solved with polynomial size* ACC *circuits, there is an* $\varepsilon > 0$ *such that* SUCCINCT 3SAT *can be solved by a nondeterministic algorithm running in* $O(2^{n-n^{\varepsilon}})$ *time, on all circuits with n inputs and* poly$(n)$ *size.*

The contrapositive says that time lower bounds can be spun into circuit lower bounds. From Theorem 2.4 it follows readily that SUCCINCT 3SAT cannot have polynomial size ACC circuits, since the consequence of Theorem 2.4 contradicts Theorem 2.3, the time lower bound for SUCCINCT3SAT.

**Corollary 2.2** SUCCINCT 3SAT *does not have polynomial size* ACC *circuits, i.e.,* NEXP $\not\subset$ ACC.

So Theorem 2.4 is now our primary target. Why might it be true? How can we spin nonuniform circuits for SUCCINCT 3SAT into a single uniform algorithm which beats $2^n$ time? Since we can allow nondeterminism in the algorithm, we could *guess* a polynomial size ACC circuit $C$ that solves SUCCINCT 3SAT, then run $C$ on our input. But how could we *check* that $C$ correctly solves SUCCINCT 3SAT? Naively, we would need to check that on all $2^n$ inputs $x$, $C(x) = 1$ iff $T(x)$ is an exponentially long satisfiable 3-CNF formula. The time lower bound (Theorem 2.3) suggests this is impossible to do in less-than-$2^n$ time.

## 3 Program Checking?

We may try draw ideas from *program checking*, a topic introduced by Blum and Kannan [BK95]. In program checking, one has a desired problem $\Pi$ in mind, and one is given a program $P$ as a black box along with an input $x$. One wishes to efficiently determine if the output of $P(x)$ equals $\Pi(x)$, i.e. if $P$ reports a correct answer on $x$, by asking questions to $P$. More formally:

**Definition 3.1** *A program checker C for a problem $\Pi$ and input x is a probabilistic polynomial time algorithm which is given black-box access to a program P and has the following properties for every P and x:*

- *If P correctly computes $\Pi$ on all inputs, then $C^P(x)$ outputs the correct answer, with high probability.*[6]

- *If $P(x) \neq \Pi(x)$ then $C^P(x)$ outputs "fail" or the correct answer, with high probability.*

What problems $\Pi$ can be checked in this way? There has been extensive work on this question; cf. the work of [GGHKR08] for a survey and recent results. SUCCINCT 3SAT doesn't seem to have a program checker, but there is another way in which SUCCINCT 3SAT can be efficiently checked. In a very influential paper, Babai, Fortnow, and Lund [BFL91] proved that every NEXP problem $\Pi$ can be recognized by a probabilistic polynomial time (PPT) algorithm with access to an arbitrary oracle which is trying to "prove" that a given instance is in $\Pi$. More precisely, for every NEXP problem $\Pi$ there is a PPT algorithm $A$ such that

- if $x \in \Pi$ then there is an oracle $O$ such that $\Pr[A^O(x) \text{ accepts}] \geq 2/3$, and

- if $x \notin \Pi$ then for all oracles $O$, $\Pr[A^O(x) \text{ rejects}] \geq 2/3$.

Informally, every $\Pi \in$ NEXP has some PPT verifier $A$ with exponentially long proofs that can be efficiently checked. Since the oracle $O$ could only be asked $\exp(|x|^k)$ different queries over all possible runs of $A^O(x)$, it follows that for every $x$ of length $n$, the corresponding oracle $O$ can be represented by an $\exp(n^k)$-bit string encoding all the possible queries and answers of $A^O(x)$. Hence this proof verification model *characterizes* NEXP. This is encapsulated by the equation NEXP = MIP. (The class MIP stands for Multiple Interactive Provers, an equivalent model to the PPT algorithm with oracle access.)

---
[6]The notation $C^P$ denotes $C$ with black-box access (i.e., oracle access) to $P$.

Let's see what happens when we try to apply the NEXP = MIP theorem directly to our situation. Recall we want to derive a SUCCINCT 3SAT algorithm that is nondeterministic and runs in less than $2^n$ time. Consider the algorithm:

> SATALG($x$):
>     Nondeterministically guess a poly($|x|$)-size circuit $C$.
>     Run a PPT algorithm $A$ for checking SUCCINCT 3SAT on $x$,
>         treating $C$ as the oracle.
>     If $A^C(x)$ accepts then *accept* else *reject*.

Unfortunately, SATALG is not a correct nondeterministic algorithm: it takes a nondeterministic guess followed by a randomized computation $A$ which could err when it rejects. So SATALG($x$) could have an accepting computation path, when $x$ is in fact a no-instance of SUCCINCT 3SAT. Moreover, it is not known how to convert arbitrary (two-sided error) PPT algorithms into efficient nondeterministic ones. (Indeed, it is open whether BPP = NEXP; that is, probabilistic polynomial time *could be as strong as* nondeterministic exponential time!) Could we possibly remove the use of randomness in the checker? The NEXP = MIP result no longer holds when you replace PPT algorithms with nondeterministic algorithms: a nondeterministic polynomial time algorithm that consults an oracle could be only as powerful as NP itself![7]

We seem to have failed to progress towards Theorem 2.4, but here's a thought. Program checking has an inherently black-box aspect: we only study the input/output behavior of a program (or proof oracle, in the case of NEXP = MIP). But our particular box of interest (SUCCINCT3SAT) is very special; by assumption, it can be modeled with a small ACC circuit. In a nondeterministic algorithm, we could guess this circuit and dissect its insides. Surely this extra information is useful.

## 4 Black Boxes Versus Circuits

What distinguishes a black box from a small circuit? If we could analyze circuits in a way which is provably better than analyzing black boxes, perhaps we could improve on what boxes can offer in the above. We can approach this improvement from two directions: try to find "easy" circuit-analysis problems, or try to find "hard" black-box-analysis problems. Or we could try both.

Consider a black box that takes $n$ inputs and prints a bit; we can query it repeatedly, and we have to determine some property of it. What is the hardest simple black-box problem? I would say that it is determining if the box will output a 1 on some input. If I want to determine this, an annoying adversary could simply answer "0" to all my queries until the last one. So this simple problem already requires $2^n$ queries to solve – a hard black-box problem.

Can we solve the problem more efficiently if we put circuits in place of black boxes? Replacing the black box with a circuit is precisely the Circuit Satisfiability problem. And if one were to define "more efficiently" to be "polynomial time" then this is the P versus NP question. We should tread lightly in this area of the jungle. Before proceeding further, let's do a sanity check on our line of thought, and assume the strongest "separation" between black-box hardness and circuit hardness. Let's assume Circuit SAT can be solved really efficiently, P = NP. Could we then prove our desired circuit lower bound? Yes.

---

[7] One can simulate a nondeterministic algorithm-with-oracle in NP, by simply guessing an accepting computation path for the algorithm (along with prospective answers for the oracle queries along the way), then checking that the oracle answers are consistent with each other. If there is no oracle that makes the nondeterministic algorithm accept, then no accepting path can exist. However, if we considered *co-nondeterministic* algorithms with oracles, then we recover NEXP again. This point will be revisited in Section 5.

**Theorem 4.1 (Karp-Lipton [KL80], attributed to Meyer)** *Suppose Circuit Satisfiability is in* P. *If* SUCCINCT *3SAT were solvable with polynomial size circuits, then all problems solvable in* $2^n$ *time would be solvable in polynomial time (therefore,* SUCCINCT *3SAT does not have polynomial size circuits, by the time hierarchy theorem).*

In fact, Meyer proves a stronger implication: if P = NP then EXP $\not\subset$ P/poly. That is, assuming Circuit Satisfiability is in P, there is an exponential time computable function that doesn't have polynomial size circuits.

The idea of the proof is to set up a fast (contradictory) simulation of every $2^n$ time algorithm $A$, assuming both P = NP and EXP $\subset$ P/poly. For simplicity let us assume $A$ is a one-tape Turing machine; the argument can be generalized for other models. On an input $x$, nondeterministically guess a polynomial size circuit $C$ that encodes the (exponentially long) computation history of $A$; that is, the truth table $T(C)$ of $C$ is a valid computation history of $A(x)$. Such a $C$ exists if EXP $\subset$ P/poly. To verify $C$ works correctly, we can *universally* (using co-nondeterminism) try all steps $i$ and all tape cells $j$, and verify that $C$ makes consistent claims about the content of cell $j$ at step $i$, by comparing the claimed content at step $i-1$ of cells $j-1$, $j$, and $j+1$. (As $A$ is a one-tape machine, the content of cell $j$ can only be affected by that of $j-1$, $j$, and $j+1$ in the previous step.) This only requires evaluating $C$ at four different pairs of indices: $(i, j)$, $(i-1, j-1)$, $(i-1, j)$, and $(i-1, j+1)$, which can be done in polynomial time. If $C$ makes consistent claims about $(i, j)$ for every $i$ and $j$, then our simulation accepts iff $C$ claims that $A(x)$ accepts. This is a $\Sigma_2$P computation, where we start with a nondeterministic guess and then universally verify our guess. But if P = NP then $\Sigma_2$P = P, so we have simulated every $2^n$ time algorithm $A$ in polynomial time, a contradiction to the time hierarchy theorem.

So the idea of using a circuit-analysis algorithm to prove a circuit lower bound has merit. But ugh... P = NP? Do we really need such a strong (probably false) algorithmic assumption? One can get away with a slower algorithm for Circuit SAT. We say that a function $f : \mathbb{N} \to \mathbb{N}$ is "half-exponential" if $f(f(n^k)^k) \leq 2^{n/2}$ for all $k > 1$. Examples of half-exponential functions are $f(n) = n^{\text{poly}(\log n)}$ and $f(n) = 2^{2^{\text{poly}(\log \log n)}}$. Carefully following the above argument, one can prove:

**Theorem 4.2 (Karp-Lipton [KL80], attributed to Meyer)** *If Circuit Satisfiability is in half-exponential time, then* EXP $\not\subset$ P/*poly.*

Now how plausible is this assumption? Unfortunately, it looks quite hard to find even a $2^{n^\varepsilon}$ time algorithm for Circuit SAT for some $\varepsilon < 1$. (Note, $2^{n^\varepsilon}$ is much larger than half-exponential.) The state of the art in satisfiability algorithms is far from half-exponential time, although steady progress has been made since Monien and Speckenmeyer [MS85]. They showed that for every $k$, there is an $\alpha_k < 1$ such that $k$-SAT is solvable in $2^{\alpha_k n}$ time, but $\lim_{k \to \infty} \alpha_k = 1$. Many improvements on the values of $\alpha_k$ have been found over the years (e.g., [Sch92, Sch02, PPSZ05, MS11, Her11]), but no one has found an algorithm for 3SAT that runs in $2^{\alpha n}$ time for every $\alpha > 0$. The *Exponential Time Hypothesis* of Impagliazzo and Paturi [IP01] states that 3-SAT (and hence, Circuit SAT) requires $2^{\alpha n}$ time for some $\alpha > 0$, and a majority of researchers believe this hypothesis.

## 5 Backtrack

We have reached an impasse, so let's review how we got here. We wanted to prove that, if SUCCINCT 3SAT can be solved in ACC, then we can design a faster-than-$2^n$ nondeterministic algorithm for SUCCINCT 3SAT (a contradiction). We started by imagining a nondeterministic algorithm which guesses a polynomial size circuit for SUCCINCT 3SAT and checks correctness of that circuit, but that seemed impossible to efficiently implement directly. Using the ideas behind NEXP = MIP, we proposed an algorithm SATALG that guesses an "oracle circuit" and verifies that, but the verification doesn't seem to be implementable nondeterministically. We concluded that the black-box nature of NEXP = MIP made it insufficient for our purposes, so we began looking for circuit-analysis problems that are easier than the corresponding black-box problems. Examining an argument of Karp-Lipton-Meyer, we found that a half-exponential algorithm for Circuit Satisfiability would imply circuit size lower bounds. But such an algorithm may not exist. Here are two observations:

1. If we assume NEXP has small circuits, then all sorts of expensive computations can be captured with small circuits. So with a nondeterministic algorithm, we could always guess more circuits encoding additional information that may help verify other circuits.

2. We have not yet used any particular properties of ACC circuits: all of our considerations would apply equally well for P/poly.

Let's focus on the first point; the second will be handled later. Impagliazzo, Kabanets, and Wigderson [IKW02] proved that if NEXP $\subset$ P/poly, then not only does SUCCINCT 3SAT have polynomial size circuits, but in fact for every circuit succinctly representing a satisfiable 3-CNF formula, there is another circuit succinctly representing a *satisfying assignment* for that formula.

Let $T(x)$ be the truth table of a string $x$, provided $x$ is encoded as a circuit. (If $x$ does not encode a valid circuit, let $T(x) = 0^{2^{|x|}}$.) For a circuit $x$, let $F_x$ be the 3-CNF formula encoded by $T(x)$. (If $T(x)$ does not encode a 3-CNF, let $F_x$ be the trivially false formula.)

**Theorem 5.1 ([IKW02])** *Suppose* NEXP $\subset$ P/poly. *Then for every* $x \in$ SUCCINCT 3SAT, *there is a circuit $W_x$ of poly$(|x|)$ size and $O(|x|)$ inputs such that $T(W_x)$ is a satisfying assignment to the formula $F_x$.*

The proof is an ingenious mixture of results on "hardness versus randomness" and good old-fashioned diagonalization; we do not have space to describe it here, but encourage the reader to take a look. It is not hard to show that if NEXP $\subset$ ACC, then these "satisfying assignment circuits" can be assumed to also be ACC:

**Corollary 5.1** *Suppose* NEXP $\subset$ ACC. *Then for every $x \in$ SUCCINCT 3SAT, there is an ACC circuit $W_x$ of poly$(|x|)$ size and $O(|x|)$ inputs such that $T(W_x)$ is a satisfying assignment to the formula $F_x$.*

**Proof.** NEXP $\subset$ ACC implies NEXP $\subset$ P/poly, so every $x \in$ SUCCINCT 3SAT has a succinct satisfying assignment represented by a circuit, $W_x$. Since P $\subset$ ACC, it follows that the CIRCUIT VALUE PROBLEM has polynomial size ACC circuits.[8] Therefore from $W_x$, there is an equivalent ACC circuit $W'_x$, obtained by plugging in an encoding of $W_x$ into the inputs of an ACC circuit for the CIRCUIT VALUE PROBLEM. □

(Notice again that we still have not used specific properties of ACC in the above proof.) Hence if NEXP were solvable with small ACC circuits, then every problem with an extremely long solution would always have *some* solution with an extremely efficient ACC representation.

This prompts the idea: rather than guessing a circuit for SUCCINCT 3SAT, or an oracle circuit that's verifiable with randomness, why not guess a circuit encoding a satisfying assignment for our given instance? Perhaps this is easier to check. We are immediately led to:

---
[8]Recall the CIRCUIT VALUE PROBLEM is: *given a circuit C and input x, does $C(x) = 1$?*

SATALG2($x$):
>   Nondeterministically guess a poly($|x|$)-size circuit $W_x$.
>   If $T(W_x)$ encodes a satisfying assignment to $F_x = T(x)$, then *accept* else *reject*.

Verifying that $W_x$ encodes a satisfying assignment to $F_x$ can be done in exponential time, by evaluating $W_x$ on all inputs, obtaining the string $T(W_x)$, evaluating $x$ on all inputs obtaining $F_x$, then checking that $T(W_x)$ satisfies $F_x$. Provided NEXP $\subset$ P/poly, SATALG2 will correctly solve SUCCINCT 3SAT, by Theorem 5.1. Now the interesting question is, *can* SATALG2 *be implemented to run in* $2^{n-\omega(\log n)}$ *time, assuming* NEXP $\subset$ P/poly*?* If yes, we will have finally contradicted Theorem 2.3, the time lower bound for SUCCINCT 3SAT.

Checking that a variable assignment satisfies a 3-CNF formula can be done using an amount of workspace that is only logarithmic in the size of the formula and assignment. Hence SATALG2 can be implemented to run in only polynomial space. Try all possible polynomial size circuits $W_x$, and for each $W_x$, run a logspace algorithm $A$ for checking satisfiability as follows: when $A$ needs a bit of $F_x$, evaluate the circuit $x$ on the appropriate index; when $A$ needs a bit of the assignment, evaluate $W_x$ on the appropriate index. This way, we do not have to hold the entire formula or assignment in memory at once, and we'll take only polynomial space. So if we could solve this *polynomial space* problem faster than $2^n$, we could solve SUCCINCT 3SAT in less than $2^n$ time, getting a contradiction.

This still looks algorithmically difficult to implement in faster than $2^n$ time; can the complexity of checking be reduced even further? To verify that an assignment satisfies a 3-CNF formula, one checks for all clauses that the assignment satisfies at least one of three literals in the clause. We can "iterate over all clauses" by feeding different inputs into the circuit $x$. We can compute the three literals of a particular clause of $F_x$ by evaluating $x$ at $O(|x|)$ inputs. We can compute the values of those three literals, under the assignment $T(W_x)$, by feeding three appropriate inputs into $W_x$. The picture of how to determine whether the $i$th clause of $F_x$ is satisfied by $T(W_x)$ looks like this:

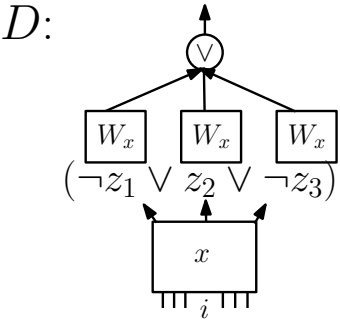

In this picture, the $i$th clause of $F_x$ is $(\neg z_1 \vee z_2 \vee \neg z_3)$, and $D(i) = 1$ iff the variable assignment encoded by $W_x$ satisfies the $i$th clause of the formula encoded by $x$. But for every $i$, $x$ can be rigged to print the $i$th clause of $F_x$, which can then be checked against $W_x$. It follows that the circuit $\neg D$ is unsatisfiable if and only if the variable assignment encoded by $W_x$ satisfies the 3CNF formula encoded by $x$. We have reduced the exponential time check of SATALG2 to Circuit Satisfiability! Let's give a revised version of our SUCCINCT 3SAT algorithm:

>   SATALG3($x$):
>   Nondeterministically guess a poly($|x|$)-size circuit $W_x$.
>   Construct the circuit $D$ made up of $x$ and $W_x$.
>   *Accept* iff $\neg D$ is unsatisfiable.

If Circuit SAT is solvable in $2^{n-\omega(\log n)}$ time for poly$(n)$-size circuits, then SATALG3 can be implemented to run in $2^{n-\omega(\log n)}$ time. But SATALG3 solves SUCCINCT 3SAT, contradicting Theorem 2.3. We have established:

**Theorem 5.2 ([Wil10])** *Assume Circuit SAT on circuits with n inputs and poly(n) size can be solved in $2^{n-\omega(\log n)}$ time. Then* NEXP *is not in* P$/$poly.

This assumption appears to be more plausible. The best known algorithms for general CNF-SAT [Sch05, DH08, CIP06] run faster than $2^{n-\omega(\log n)}$; there are even AC$^0$-SAT algorithms that beat $2^{n-\omega(\log n)}$ [CIP06, IMP11]. However it does not look easy to generalize these algorithms to unrestricted polynomial-size circuits.

## 6 Enter ACC

We are now ready to think about how to incorporate ACC into our arguments. We have found that faster Circuit SAT (an algorithmic upper bound) implies SUCCINCT 3SAT is not in P$/$poly (a circuit lower bound). Informally, this is because "SUCCINCT 3SAT has small circuits" implies that we can guess small representations of exponentially long information, and a faster Circuit SAT algorithm can help verify the correctness of the small representations. Together, the two result in a faster nondeterministic algorithm for SUCCINCT 3SAT, contradicting a known time lower bound.

Ideally, one would hope that this upper bound / lower bound connection can be extended to other circuit classes, not just P$/$poly. For each circuit class $\mathcal{C}$, we may define a corresponding $\mathcal{C}$-SAT problem: *given a generic circuit from the class $\mathcal{C}$, is it satisfiable?* As mentioned above, very little is known about the worst-case time complexity of this problem.

If we could design a faster-than-$2^n$ algorithm for $\mathcal{C}$-SAT, that should intuitively help prove a lower bound against circuits from $\mathcal{C}$: we have determined a property of circuits from $\mathcal{C}$ that is quantitatively easier than the corresponding property for black boxes.

### 6.1 Spinning restricted Circuit SAT into restricted circuit lower bounds

What goes wrong in SATALG3 when we assume that only ACC-SAT can be solved in less than $2^n$ time? Applying Corollary 5.1, if NEXP $\subset$ ACC then every satisfiable SUCCINCT 3SAT instance $x$ (construed as a circuit) has a polynomial size ACC circuit $W'_x$ such that $T(W'_x)$ is a satisfying assignment for $F_x = T(x)$, the formula encoded by $x$. This means we could guess an ACC circuit $W'_x$ instead of $W_x$ in SATALG3. But what about the circuit $x$ itself? There is no restriction on $x$, because the definition of SUCCINCT 3SAT lets $x$ be arbitrary. So the resulting circuit $D$ that we produce to check $x$ and $W'_x$ will be unrestricted as well. Hence an ACC-SAT algorithm won't necessarily run correctly on $D$.

However, assuming P $\subset$ ACC, Corollary 5.1 tells us that for *every* polysize circuit $x$, there *exists* an equivalent, polynomial size ACC circuit $x'$. Again, this is because the CIRCUIT VALUE PROBLEM is in ACC, hence we can simulate the behavior of all unrestricted circuits using ACC circuits. So we could try to guess this ACC circuit $x'$, and use that in place of $x$ in the construction of the circuit $D$. Then, the circuit $D$ will have an ACC circuit $x'$ composed with three copies of an ACC circuit $W'_x$, which will altogether be an ACC circuit. That is, we are proposing the following modification to SATALG3:

> SATALG4($x$):
>     Nondeterministically guess a poly($|x|$)-size ACC circuit $W'_x$.
>     Nondeterministically guess a poly($|x|$)-size ACC circuit $x'$.
>     Verify that $x$ and $x'$ are equivalent **(???)**
>     Construct the ACC circuit $D$ made up of $x'$ and $W'$.
>     *Accept* iff $\neg D$ is unsatisfiable.

SATALG4 now checks the satisfiability of an ACC circuit, rather than an unrestricted circuit. By an argument analogous to what we gave for SATALG3, a $2^{n-\omega(\log n)}$ algorithm for ACC Circuit SAT for $n$-input poly($n$)-size circuits would appear to give our desired nondeterministic algorithm for SUCCINCT 3SAT.

However, as the **(???)** indicates, there remains a hole to be filled in. We have to verify that the input $x$ and our guess $x'$ are really computing the same function. Can that be done with a faster ACC-SAT algorithm? The usual way of checking equivalence of $x$ and $x'$ would be to set up a Circuit SAT instance of the form

$$E(i) = (x(i) \vee x'(i)) \wedge (\neg x(i) \vee \neg x'(i)),$$

and check if $E$ is satisfiable. But this $E$ contains a copy of the unrestricted circuit $x$, so $E$ is also unrestricted! It seems hopeless to turn $E$ into an ACC satisfiability question. We could guess an equivalent ACC circuit $E'$, but that wouldn't seem to help; we'd then have to verify that $E$ is equivalent to $E'$, and $E, E'$ are only larger than $x, x'$.

This is an annoying and impossible-looking problem. The key to solving it is to *use the assumption that* NEXP *has small circuits, and guess small "helper" circuits in the* SUCCINCT 3SAT *algorithm*. If NEXP $\subset$ ACC then many types of functionality could be guessed in ACC form; if we choose the right functionality, our ACC circuit SAT algorithm can help verify the functionality. We must also avoid an infinite regress: eventually we must have some guessed circuits that can be directly checked for correctness.

What else can we guess? We want to obtain an ACC $x'$ that's provably equivalent to the input $x$, in the sense that both circuits produce the same outputs. But the output of a circuit is only one bit of information. Why not guess an ACC circuit that captures even more information about $x$? When we construe $x$ as a circuit and evaluate it on input $i$, many bits of information are produced: on an input $i$, bit values are carried along every wire in $x$. Assuming P $\subset$ ACC, these bits can be produced in ACC: there are poly($n$)-size ACC circuits $C$ which take as input an (unrestricted) circuit $x$ described in $n$ bits, an input $i$ to $x$, and an integer $j$, such that

$C(x, i, j)$ prints the value output by the $j$th gate of the circuit $x$, when $x$ is evaluated on input $i$.

(Determining this value can be done in polynomial time given $(x, i, j)$, so if P $\subset$ ACC then there are ACC circuits that can determine the value.) Provided we have a circuit $C$ meeting the above specification, then for every $i$ we have $x(i) = C(x, i, j^\star)$, where $j^\star$ is the index of the output gate of $x$. By setting $x' = C(x, \cdot, j^\star)$, we have an ACC circuit equivalent to $x$.

Supposing we guess this circuit $C$, we have to verify it is correct on our input $x$. At this point, it appears we have made our job only harder, since $C$ takes strictly more inputs than our original guess $x'$ did! But by forcing the guessed circuit to print *more* valid information about $x$, we can more easily verify that *all* the information is correct.[9] For instance, if we find an AND gate $j$ in $x$ where the values output by $C(x, i, \cdot)$ imply that $j$ receives the inputs 1 and 0, but $C(x,i,j) = 0$, then we have detected an error in $C$. Conversely, if $C(x,i,\cdot)$ manages to make consistent claims about the inputs and outputs of every gate of $x$ on input $i$ (that is, OR gates always output the OR of their inputs, ANDs always output the AND of their inputs, NOTs always negate), then we know that $C$'s claim about the final output of $x(i)$ must also be correct.

---

[9]There is a similar principle behind error-correcting codes: by appending a message with more information about the message, one can still verify the content of the original message if some bits get flipped. The principle can also be seen in the technique of *algebrization*, where in order to better manipulate a Boolean function $f(x_1, \ldots, x_n)$, one "lifts" $f$ to a low-degree multivariate polynomial $p$ which is equivalent to $f$ on the set $\{0,1\}^n$. Querying $p$ on points *outside* of the set $\{0,1\}^n$, one can often gain considerable advantages in verifying and manipulating $f$.

We can think of $C$ as encoding a *satisfying assignment* to an exponentially large constraint satisfaction problem: for every input $i$ to the circuit $x$ and every gate $j$ of $x$, the $(i,j)$ constraint is that $C$'s claimed inputs to gate $j$ in the evaluation of $x(i)$ are consistent with $C$'s claimed output of gate $j$. This constraint satisfaction problem has a succinct description – namely, the circuit $x$ itself. Every circuit $x$ with $s$ gates can be represented as a set of tuples

$$S_x = \{\langle j, j_1, j_2, g\rangle \mid j = 1,\ldots,s; j_1, j_2 < j; g \in \{\text{AND, OR, NOT, INPUT}\}\}.$$

The tuple $\langle j, j_1, j_2, g\rangle$ says that the $j$th gate of $x$ takes its inputs from the output of gate $j_1$, the output of gate $j_2$, and $j$ has gate type $g$.[10] (For $j = 1,\ldots,n$, we use the convention that gate $j$ corresponds to the $j$th bit of input, so the integers $j_1$ and $j_2$ equal 0, and $g = \text{INPUT}$.) From the set $S_x$, we can define a function $G_x$ which takes $j$ as input and prints the rest of the tuple $\langle j_1, j_2, g\rangle$ from $S_x$. Since the number of all possible inputs to $G_x$ is only $s$ (the number of gates in $x$), $G_x$ can be implemented in Boolean logic with an $O(\log|x|)$-size collection of CNF formulas, each with $O(\log|x|)$ variables and $O(|x|)$ clauses.

So to check that a guessed ACC circuit $C$ is correct, we can use the following ACC circuit:

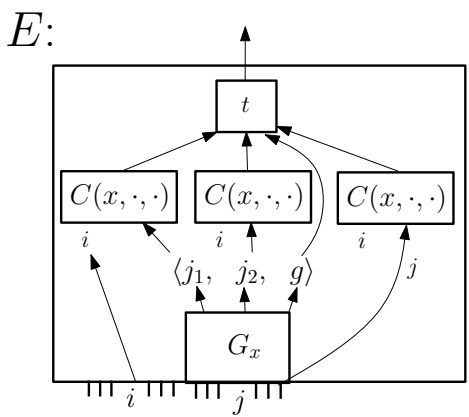

The $O(1)$-size circuit $t$ takes bits $b_1, b_2$ (from $j_1, j_2$) a bit $b$ (from $j$), and a gate type $g$; $t(b_1, b_2, b, g) = 1$ iff $g(b_1, b_2) = b$. For example, $t(1,0,1,OR) = 1$, since $OR(1,0) = 1$.

For a given $i$, $E(i,j) = 1$ for all $j$ iff $C$ outputs the correct value of every wire in $x(i)$. Now define

$$E'(i) = \bigwedge_{j=1}^{n} [C(x,i,j) \Leftrightarrow x_j] \wedge \bigwedge_{j=n+1}^{s} E(i,j).$$

(The first group of ANDs check that $C$ represents the input gates correctly.) The circuit $E'$ is also ACC, has exactly the same number of inputs as the circuit $x$, and $\neg E'$ is unsatisfiable iff $C$ is correct on all inputs $i$ to $x$. Our modified algorithm now looks like:

---

[10]Without loss of generality, we may assume every gate of the circuit has at most two inputs.

> SATALG5($x$):
>     Nondeterministically guess a poly($|x|$)-size ACC circuit $W'$.
>     Nondeterministically guess a poly($|x|$)-size ACC circuit $C$.
>     Construct the collection of CNFs representing $G_x$.
>     Construct the ACC circuit $\neg E'$ made up of $C$ and $G_x$.
>     If $\neg E'$ is satisfiable then *reject*
>     /⋆ *at this point, C must be correct* ⋆/
>     Define $x' = C(x, \cdot, j^\star)$ where $j^\star$ corresponds to the output gate of $x$.
>     Construct the ACC circuit $D$ made up of $x'$ and $W'$.
>     *Accept* iff $\neg D$ is unsatisfiable.

We have finally reached a correct algorithm for SUCCINCT 3SAT (under the assumption that NEXP $\subset$ ACC) with the property that if ACC satisfiability can be solved faster than $2^n$, then SATALG5 can be implemented to run faster than $2^n$, yielding our desired lower bound.

**Theorem 6.1 ([Wil11])** *If* ACC *Circuit SAT can be solved on circuits with n inputs and $n^k$ size in $2^{n-\omega(\log n)}$ time for every k, then* NEXP *is not contained in* ACC.

We only needed a few basic closure properties of ACC in the above argument: if you take a polynomially large AND of different ACC circuits, then the result is still an ACC circuit; given two ACC circuit families $\{C_n\}$ and $\{D_n\}$, if you define a circuit family $\{E_n\}$ by the rule

$$E_n(x_1, \ldots, x_n) = C_{n^k}(D_n(x_1, \ldots, x_n), \ldots, D_n(x_1, \ldots, x_n))$$

for some fixed $k$, this "composition" of $\{C_n\}$ and $\{D_n\}$ is also an ACC circuit family. Most well-studied circuit classes satisfy these composition properties, so the above considerations apply to them as well: faster circuit satisfiability for a restricted class $\mathcal{C}$ entails lower bounds for solving problems in $\mathcal{C}$. Intuitively, the difficulty faced by researchers who design fast algorithms for verification of certain kinds of circuits is related to the difficulty of proving that certain problems *can't* be efficiently solved with these kinds of circuits.

## 6.2 ACC Circuit Satisfiability

It remains to prove that ACC circuit satisfiability really does have a faster algorithm. We can discover this algorithm by studying a known decomposition result for ACC circuits, from work initiated by Yao [Yao90], continued by Beigel and Tarui [BT94], Allender and Gore [AG94], and Green *et al.* [GKRST95]. The decomposition result says that every ACC circuit family can be expressed as a family of functions

$$\{g_n(h_n(x_1, \ldots, x_n))\},$$

where $h_n$ is a "sparse" multilinear polynomial, and $g_n$ is a "sparse" lookup table.

**Lemma 6.1 ([Yao90, BT94, AG94])** *There is an algorithm and function $f : \mathbb{N} \times \mathbb{N} \to \mathbb{N}$ such that given an* ACC *circuit $C$ with* $\text{MOD}_m$ *gates of n inputs, depth d, and size s, the algorithm outputs a function $g : \{0, \ldots, K\} \to \{0, 1\}$ and a multilinear polynomial $h(x_1, \ldots, x_n)$ with K monomials, such that $C \equiv g \circ h$, where $K = 2^{O(\log^{f(d,m)} s)}$. The algorithm takes at most $\tilde{O}(K)$ time.*

Call this transformation the *polynomial decomposition* for ACC. The function $f(d,m)$ is estimated to be no more than $m^{O(d)}$. The high-level idea behind the decomposition is to first convert every OR and AND gate in the ACC circuit to low-degree polynomials (i.e., low fan-in ANDs of MOD2 gates) using randomness, "push" these low fan-in ANDs down to the bottom of the circuit, derandomize the construction using pairwise independence and a MAJORITY gate at the top, then use more sophisticated polynomial tricks to "push" the remaining layers of MOD gates into the top gate, which remains a symmetric function throughout the transformation. At the end, what remains is a symmetric function of a quasipolynomial number of ANDs, which can be represented by a $g$ of $h$ in the above manner. Of course this is a very rough description; the reader should check the references for more details.

What does this polynomial decomposition algorithm say about solving satisfiability for ACC? Razborov and Smolensky's lower bounds on $AC^0[p]$ can be seen as "approximations by polynomials" – they show that small $AC^0[p]$ circuits can be approximated on many points by low-degree polynomials, so limitations on representing functions with low-degree polynomials can be ported over to limitations on $AC^0[p]$. Lemma 6.1 allows the polynomial $h$ to output a number of possible values; those values are then filtered down to a single bit by another function $g$. While we may not approximate an ACC circuit very well with a polynomial, we can still simulate a great deal of the *computation* in an ACC circuit with a polynomial: after the evaluation of the polynomial $h$, we are only a $g$-evaluation away from the ACC circuit's output. (In fact, Green *et al.* [GKRST95] prove that $g$ can be made a specific, simple function: the "middle bit" function.)

Polynomials are nice, but what good do they serve for satisfiability algorithms? The short answer is: *the Fast Fourier Transform*. Less ambiguously, if we are given a multilinear polynomial in its coefficient representation (we are told the coefficients of the $2^n$ possible monomials), then we can determine that polynomial's value on all points in $\{0,1\}^n$, in only $O(2^n \cdot \text{poly}(n))$ time. That is, from the coefficient representation of the polynomial we can quickly compute the point representation. This is very nice; we are spending only $\text{poly}(n)$ time per evaluation point, even though our original polynomial could have been arbitrary – it could have $2^n$ different coefficients!

There are several ways to derive a COEFFICIENT-TO-POINT algorithm. Perhaps the most natural one is a recursive strategy. We are given a multilinear polynomial $p(x_1,\ldots,x_n)$, and wish to compute a table $T$ of $2^n$ entries such that

$$T = [p(0,\ldots,0,0), p(0,\ldots,0,1), \ldots\ldots, p(1,\ldots,1,0), p(1,\ldots,1,1)].$$

If $n=1$, we can return $T = [p(0), p(1)]$ in unit time. When $n > 1$, because $p$ is multilinear we can write it as

$$p(x_1,\ldots,x_n) = x_1 q_1(x_2,\ldots,x_n) + q_2(x_2,\ldots,x_n).$$

That is, we can split $p$ into sums of monomials which include $x_1$, and sums of monomials which do not include $x_1$. Recursively calling our algorithm on $q_1$ and $q_2$, we receive two tables $T_1$ and $T_2$ of $2^{n-1}$ numbers each. Notice that $p(0, x_2,\ldots,x_n) = q_2(x_2,\ldots,x_n)$, and $p(1, x_2,\ldots,x_n) = q_1(x_2,\ldots,x_n) + q_2(x_2,\ldots,x_n)$. Therefore the corresponding $2^n$ size table for $p$ is

$$T = \left[T_2[1],\ldots,T_2[2^{n-1}], T_1[1]+T_2[1],\ldots,T_1[2^{n-1}]+T_2[2^{n-1}]\right].$$

The merging of tables can be done in $O(2^n \cdot \text{poly}(n))$ time, so the running time recurrence is

$$R(2^n) = 2 \cdot R(2^{n-1}) + O(2^n \cdot \text{poly}(n)),$$

which solves to $O(2^n \cdot \text{poly}(n))$.[11]

---

[11] Here we are assuming that the sizes of coefficients in the polynomial $p$ are negligible, so the bit-complexity of arithmetic does not play a significant role in the running time. This assumption is valid for the polynomials we are considering.

How can a COEFFICIENT-TO-POINT algorithm lead to a faster ACC SAT algorithm? There seem to be two sticking points.

1. The above works directly on a multilinear polynomial $h$, but we need an algorithm that works for a $g$ of an $h$.

2. The above runs in $2^n$ time, but we need a SAT algorithm that runs faster than $2^n$.

Addressing the first point is straightforward. We can evaluate $h$ on all $2^n$ points, and after we have produced the $2^n$ table, we can determine all the distinct numbers in the table and check if some number makes $g$ output 1. Since $h$ is a "sparse" polynomial, the total number of different numbers in $T$ is "sparse" so this can be done in no more than $O(2^n \cdot \text{poly}(n))$ time.

The second point looks more difficult to overcome. To apply COEFFICIENT-TO-POINT and get less-than-$2^n$ time, we need to work with a polynomial that has fewer than $n$ variables. This would seem to require that our original ACC circuit has fewer than $n$ inputs – something we are not willing to concede.

There is a trick to circumvent this problem, and it exploits two observations. First, note the circuit satisfiability problem amounts to asking if the OR of some $2^n$ circuits (with no free variables) evaluates to 1. Second, if we take an OR of many copies of an ACC circuit of depth $d$, the result is an ACC circuit of depth $d+1$, because ACC circuits allow for OR gates of unbounded fan-in.

Suppose we take a subset of $k$ of the $n$ inputs to an ACC circuit $C$, evaluate $C$ on all $2^k$ possible values of this subset, then take the OR of these $2^k$ circuit copies induced by the different evaluations. The resulting circuit $C'$ has the properties:

- $C'$ has only $n-k$ free inputs.
- If $C$ had size $s$, then $C'$ has size $O(2^k \cdot s)$.
- $C'$ is still an ACC circuit (but with one more level of depth).
- $C$ is satisfiable iff $C'$ is satisfiable.

Call this transformation the *k-blowup* of the circuit $C$. Basically, we have "brute-forced" the SAT problem for $C$ on a $k$-subset of the inputs to $C$. This blows up the size, but it *decreases the number of input variables*, something we are interested in doing, but with polynomials. However, because $C'$ is an ACC circuit, we can still perform a polynomial decomposition on the circuit $C'$, then work with the underlying $(n-k)$-variate polynomial.

Now we are ready to stitch together the ACC satisfiability algorithm, which is given a circuit $C$ with $n$ inputs and $s$ size.

---
ACCSAT($C$):
  Let $k = n^{1/(2f(d,m))}$.
  Compute $C'$, the $k$-blowup of $C$, which has size $\leq 2^k s$.
  Decompose $C'$ into $g \circ h$,
    where $h$ has $n-k$ variables and $K = 2^{O(k^{f(d,m)} + \log^{f(d,m)} s)}$ monomials.
  Evaluate $h$ on all $2^{n-k}$ points in $O(2^{n-k} \text{poly}(n) + K)$ time.
  Output *satisfiable* iff $g \circ h$ equals 1 on at least one point.

---

When $s \leq 2^{n^{o(1)}}$, we have $K = 2^{n^{1/2+o(1)}}$ and ACCSAT runs in about $2^{n-n^{1/(2f(d,m))}}$ time.

**Theorem 6.2 ([Wil11])** *ACC Circuit Satisfiability for subexponential size circuits can be computed in $2^{n-n^\varepsilon}$ time, for some $\varepsilon > 0$ which depends on the depth and modulo gates of the input circuit.*

Combining this with Theorem 6.1, we conclude that SUCCINCT 3SAT is not in ACC.

## 7 Further Directions

There are two obvious directions to continue in:

- **Find an easier problem that is not in** ACC. It is possible the ideas here may be extended to find an EXP problem (not just NEXP) which isn't in ACC. More precisely, faster $\mathcal{C}$-SAT for a circuit class $\mathcal{C}$ ought to lead to EXP $\not\subset \mathcal{C}$. Here is my extremely hand-wavy argument for this. Intuitively, a faster $\mathcal{C}$-SAT algorithm reveals a *weakness* in representing computations with $\mathcal{C}$ circuits. The class $\mathcal{C}$ is *not* like a set of black boxes: these circuits cannot hide a satisfying input so easily. Moreover, a faster SAT algorithm for $\mathcal{C}$ highlights a *strength* of algorithms that run in less-than-$2^n$ time: they can solve nontrivial satisfiability problems on circuits from $\mathcal{C}$. That is, my intuition is that a faster $\mathcal{C}$-SAT algorithm shows "less-than-$2^n$ algorithms are strong" and "$\mathcal{C}$-circuits are weak" – so perhaps $2^n$ time can be separated from $\mathcal{C}$-circuits using satisfiability algorithms alone.

- **Prove stronger circuit lower bounds for** SUCCINCT 3SAT. In order to separate NEXP from a circuit class $\mathcal{C}$, we need only design faster satisfiability algorithms for $\mathcal{C}$-circuits. In fact, it suffices to find a faster algorithm for the problem: *given a circuit $C \in \mathcal{C}$ where you are promised that either $C$ is unsatisfiable or $C$ accepts $1/2$ of its inputs, determine which is the case.*[12] Hence we only need to derandomize certain promise problems to establish the lower bounds. So far I have personally found it convenient to think about satisfiability directly, but eventually we will probably find the promise problem to be an easier chore.

There are several other not-so-obvious directions. One possibility is to prove almost-everywhere ACC circuit lower bounds. Right now we can only say that ACC circuits can't solve SUCCINCT 3SAT on infinitely many input lengths. But we don't believe that some input lengths are inherently easier than others, so our lower bound ought to be extendable to all but finitely many input lengths.

Another angle is to try proving new separations of uniform complexity classes. Can we prove NP is not equal to *uniform* ACC, where a single efficient algorithm given input $0^n$ can construct the $n$th circuit in the family? Can some of the ideas here be used to finally prove NEXP $\neq$ BPP?

Finally, much of our analysis centered around specific properties of SUCCINCT 3SAT. Might it be the case that other "Succinct" problems are useful for lower bounds, too?

**Acknowledgments.** I thank Anup Rao for an inspiring discussion, and Virginia for her patience while I was finishing this article (we were supposed to be touring Barcelona, not circuit complexity).

---

[12]Goldreich and Meir have observed that this consequence follows readily from combining results of the two papers [Wil10, Wil11].